\documentclass[useAMS,usenatbib]{mn2e}
\usepackage{graphicx}
\usepackage{amsmath}
\usepackage{bm}
\usepackage{epsfig}
\usepackage{amssymb}
\usepackage{natbib}
\usepackage{threeparttable} 
\usepackage{subfigure}
\usepackage{setspace}
\usepackage{multirow}    
\usepackage{appendix}
\def\simmore{\mathbin{\lower 3pt\hbox
     {$\rlap{\raise 5pt\hbox{$\char'076$}}\mathchar"7218$}}}   %> or of order

\title[Investigate mHz QPO in 4U 1636$-$53]{Discovery of a correlation between the frequency of the mHz quasi-periodic oscillations and the neutron-star temperature in the low-mass X-ray binary 4U 1636--53}
\author[Ming Lyu et al.]
{Ming Lyu$^1$\thanks{E-mail: m.lyu@astro.rug.nl}, Mariano M\'endez$^1$ and Diego Altamirano$^2$ \\
$^1$Kapteyn Astronomical Institute, University of Groningen, PO BOX 800, NL-9700 AV Groningen, the Netherlands\\
$^2$School of Physics and Astronomy, University of Southampton, Southampton, SO17 1BJ, UK
}

\begin{document}

\date{Accepted XXXX. Received XXXX; in original form XXXX}

\maketitle

\label{firstpage}

\begin{abstract}
We detected millihertz quasi-periodic oscillations (QPOs) in an XMM-Newton observation of the neutron-star low-mass X-ray binary 4U 1636$-$53. These QPOs have been interpreted as marginally-stable burning on the neutron-star surface. At the beginning of the observation the QPO was at around 8 mHz, together with a possible second harmonic. About 12 ks into the observation a type I X-ray burst occurred and the QPO disappeared; the QPO reappeared $\sim 25$ ks after the burst and it was present until the end of the observation. We divided the observation into four segments to study the evolution of the spectral properties of the source during intervals with and without mHz QPO. We find that the temperature of the neutron-star surface increases from the QPO segment to the non-QPO segment, and vice versa. We also find a strong correlation between the frequency of the mHz QPO and the temperature of a black-body component in the energy spectrum representing the temperature of neutron-star surface. Our results are consistent with previous results that the frequency of the mHz QPO depends on the variation of the heat flux from the neutron star crust, and therefore supports the suggestion that the observed QPO frequency drifts could be caused by the cooling of deeper layers.

\end{abstract}


\section{introduction}
 A class of quasi-periodic oscillations (QPOs) at frequencies of a few mHz were first detected by \citet{revni01} in three neutron-star low-mass X-ray binaries (LMXBs), 4U 1608--52, 4U 1636--53, and Aql X-1. The low frequency range ($7-9$ mHz) spanned by these QPOs, and the strong flux variations at low photon energies ($<$ 5 keV), made them different from other QPOs found in neutron star systems, e.g., the kilohertz (kHz) QPOs \citep[see, e.g.,][]{van00,Mendez00a,Mendez01,belloni05,Linares05,Jonker05, Boutloukos06,mendez06,Vanderklis06, Altamirano08d,sanna10} and Low-frequency QPOs \citep[see, e.g.,][]{Wijnands99a,Psaltis99b,Belloni02,Straaten02,Straaten03,van04,Altamirano05, diego08b,diego12}. The mHz QPOs appear only when the source covers a particular range of X-ray luminosities, $L_{\rm 2-20 keV} \simeq (5-11) \times 10^{36}$ ergs s$^{-1}$, and the QPOs become undetectable after a type I X-ray burst \citep{revni01,diego08}. Those unique observational features suggest that the mechanism responsible for the mHz QPOs is different from the one that produces the other QPOs.
 
\citet{revni01} proposed that the mHz QPOs were due to a special mode of nuclear burning on the neutron star surface, which only occurs within a certain range of mass accretion rate. \citet{yu02} found that in 4U 1608--52 the frequency of the kHz QPO was anti-correlated with the $2-5$ keV X-ray count rate associated with a 7.5 mHz QPO present in the same observation. This result further supported the nuclear burning interpretation of the mHz QPOs: The inner disc is slightly pushed outward in each mHz QPO cycle by the stresses of radiation coming from the neutron star surface as the luminosity increases. 
 
\citet{heger07} proposed that the mHz QPOs are due to marginally stable nuclear burning of Helium on the surface of accreting neutron stars. The characteristic timescale of the oscillations in this model is close to the geometric mean of the thermal and accretion timescales for the burning layer, $(t_{therm}\times t_{accr})^{1/2}$, about 100 seconds, remarkably consistent with the 2 minutes period of the mHz QPOs. Notwithstanding, the marginally stable burning in the model of \citet{heger07} occurs only within a narrow range of mass accretion rate, close to the Eddington rate, one order of magnitude higher than the value implied by the average X-ray luminosity. If the model is correct, the local accretion rate at the burning depth can be higher than the global accretion rate.

 \citet{diego08} found that the frequency of the mHz QPO in 4U 1636--53 systematically decreases with time until the QPO disappears before a type I X-ray burst. This behaviour further supported the idea that the mHz QPOs are closely related to nuclear burning on the neutron star surface. In addition, the work by \citet{diego08} offered a way to predict the occurrence of X-ray bursts by measuring the frequency of the mHz QPOs. \citet{linares10} detected a mHz QPO at a frequency of about 4.5 mHz in the neutron star transient source IGR J17480-2446 in the globular cluster Terzan 5. The persistent luminosity of this source when mHz QPOs were observed was $L_{2-50 keV}$ $\sim 10^{38}$ erg s$^{-1}$, about an order of magnitude higher than that observed in previous mHz QPO sources. \citet{linares12} found that in IGR J17480-2446 the thermonuclear bursts smoothly evolved into a mHz QPO when accretion rate increased, and vice versa. This evolution is predicted by the one-zone models and simulations of the marginally stable burning by \citet{heger07}, further supporting the idea that the mHz QPO in IGR J17480-2446 is due to marginally stable burning on the neutron star surface. 
 
\citet{keek09} showed that turbulent chemical mixing of the accreted fuel combined with a higher heat flux from the crust is able to explain the observed critical accretion rate at which mHz QPOs are observed, and that the frequency drift before X-ray bursts could be due to the cooling of the deep layers where the quasi-stable burning takes place. Furthermore, \citet{keek14} investigated how the transition between unstable and stable nuclear burning is influenced by the composition of the accreted material and nuclear reaction rates, and concluded that no allowed variation in accretion composition and reaction rate is able to generate a transition between burning regimes at the observed accretion rate.

4U 1636--53 is a neutron-star LMXB with a 0.1-0.25 M$_{\odot}$ star companion, located at a distance of about 6 kpc \citep{Giles02,galloway06}. Its orbital period is about 3.8 hr \citep{pedersen82}, and the system harbours a neutron star with a spin frequency of 581 Hz \citep{zhang97,stro02}. In addition to the mHz QPOs, the source shows all kinds of X-ray bursts \citep[e.g.][]{galloway08,zhang11} and also the full range of spectral states of the persistent emission \citep{Disalvo03,belloni07,diego08b}, making it an excellent candidate to explore the relation between the mHz QPOs and X-ray bursts, and also the spectral properties of the source in the mHz QPO cycle. 
  
Since they are likely related to nuclear burning on the neutron star surface, the spectral and timing analysis of the mHz QPOs provides an insight into the their origin and the process of marginally stable nuclear burning on the accreting neutron star surface. In this work, we investigate the frequency behaviour of the mHz QPO before an X-ray burst, and the reappearance of the QPO after the burst using one XMM observation. Thanks to the low energy coverage of the XMM data, this is the first time that the mHz QPOs have been investigated below $\sim$ 3 keV, and down to $\sim$ 0.8 keV. We also study the properties of the source spectrum in different segments with and without mHz QPO before and after the X-ray burst. 
  
\section{observation and data reduction}
 We used an XMM-Newton observation of 4U 1636--53 (ObsID 0606070301) performed on September 5, 2009. The data were collected with the European Photon Imaging Camera, EPIC-PN, using the Timing mode, with a total duration of 43.2 ks. The EPIC cameras cover the 0.15-12 keV energy band with a time resolution of 0.03 ms in the timing mode \citep{xmm01}.

We reduced the data using the Science Analysis System (SAS) version 13.5.0, with the latest calibration files applied. We fixed some time jumps in the raw event file due to a duplicated packet in the first extension of the EPIC-PN AUX ODF file, following the recommendation of the XMM-Newton EPIC Calibration group. We applied the Rate-Dependent PHA (RDPHA) correction via the command {\tt epproc} to account for the energy scale rate dependence in EPIC-PN Timing mode exposures when producing the calibrated event file. Following the recommendation of the EPIC calibration group, we did not apply the {\tt epfast} correction, since this one was superseded by the RDPHA correction. We used the command {\tt barycen} to convert the arrival time of each photon to the barycenter of the solar system. We found that there was moderate pile up in the observation, and therefore we excluded the central two columns (RAWX=[37,39]) of the PN CCD in the analysis to mitigate these effects. For all light curves and spectra in this work we only selected single and double events ({\sc pattern} $\leq $ 4) for extraction. 

\subsection{Spectral data}

\begin{figure*}
%\center
\includegraphics[height=0.45\textwidth]{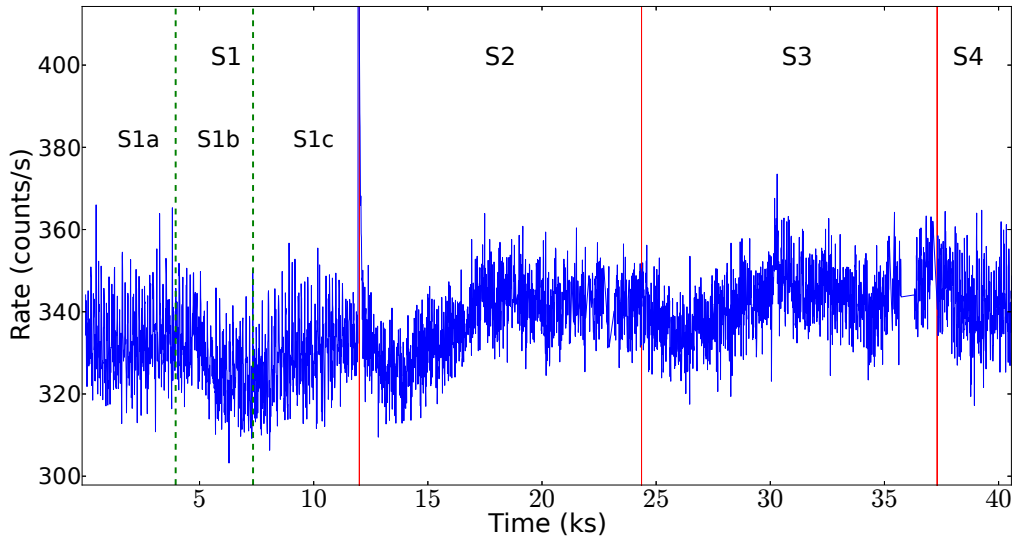}
\caption{Light curve of 4U 1636--53 with XMM-Newton PN in the $0.8-11$ keV range. The time resolution is 10 seconds. Each segment is marked with the name (S1-S4) in the plot; the red vertical lines indicate the borders between the segments. The two green dashed lines show the boundaries between the three subsegments (S1a-S1c) in the first segment. A type I X-ray burst happens at about 12 ks from the start of the observation.}
\label{lc}
\end{figure*}

 We excluded the events at the edge of the CCD and next to a bad pixel ({\sc flag} $=$ 0) to produce the spectra. We generated the redistribution matrix file (RMF) and the ancillary response file (ARF) using {\tt rmfgen} and {\tt arfgen}, respectively, for the latter using the extended point spread function (PSF) model to calculate the encircled energy correction. Since the whole region of the CCD was contaminated by source photons, we extracted the background spectrum from another timing mode observation of the black-hole candidate GX 339--4 (ObsID 0085680601) when the source was in quiescence, on the basis of similar sky coordinates and column density along the line of sight (For more details on the methodology used for the extraction of the background spectrum, please refer to \citet{hie11} and \citet{sanna13}). Finally, we rebinned the spectra to have at least 25 counts per background-subtracted channel, with an oversampling of the energy resolution of the PN detector of a factor of 3, and fitted all spectrums in 0.8-11 keV range.   
  
\subsection{Timing data}  

\begin{figure}
\center
\includegraphics[height=0.48\textwidth,angle=-90]{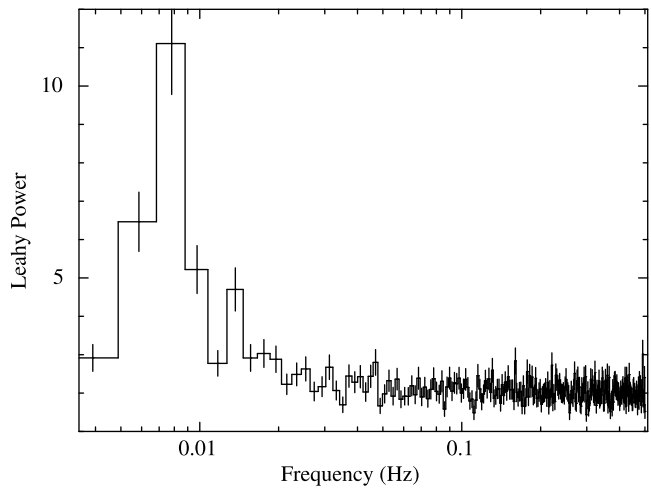}
\caption{Average power spectrum of the total observation of 4U 1636--53 in the $0.8-11$ keV range, calculated from the PN light curve at 1-second resolution. The power spectrum was rebinned logarithmically for clarity. A significant QPO at about 7$-$8 mHz and its harmonic signal are apparent in the plot.}
\label{ps}
\end{figure}

We generated a 1-s resolution light curve (we show a 10-s resolution light curve in Figure \ref{lc} for clarity) and an average power spectrum (Figure \ref{ps}) of the whole observation in the 0.8-11 keV range, after excluding instrument dropouts and an X-ray burst that took place $\sim$ 12 ks from the start of the observation. To produce the average power spectrum of the observation, we calculated the Fourier transform of intervals of 512-s duration using the command {\tt sitar$\_$avg$\_$psd} in the ISIS Version 1.6.2-27 \citep{houck2000}, and rebinned the average power density spectrum logarithmically via the command {\tt sitar$\_$lbin$\_$psd}. The frequency range of the power spectrum is from 1.95$\times$10$^{-3}$ to 0.5 Hz. A strong QPO at about 7$-$8 mHz and its second harmonic are apparent in the power spectrum (see Figure \ref{ps}). 

\begin{table*}
\caption{Information for the four segments in which we divided the XMM-Newton observation of 4U 1636$-$53.}
\begin{tabular}{|c|c|c|c|c|}
\hline
\hline
Segment number   &          Length (ks)    &   Exposure time (ks)$^{1}$ & Average count rate (counts s$^{-1}$)$^{2}$ & mHz QPO     \\
\hline
   S1         &                12      &       11.69    & 331$\pm 20$  &   Yes  \\
   S2         &                12.36   &       11.38    & 338$\pm 20$  &    No  \\
   S3         &                12.95   &       11.92    & 342$\pm 19$  &    No  \\                                                                                                                                                                                                                 
   S4         &                3.49    &       3.427    & 342$\pm 20$  &   Yes   \\
\hline
\end{tabular}                                                                                                                              
\medskip
\\
(1) The final exposure time excludes X-ray bursts, background flares and instrument dropouts. (2) Here we give the standard deviation of the count rate in each segment.
\label{overall}
\end{table*}               

\begin{figure}
\center
\includegraphics[height=0.45\textwidth]{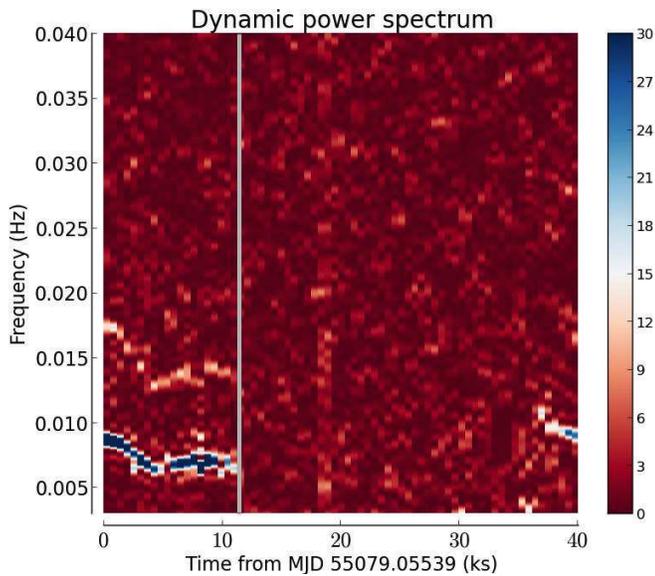}
\caption{Dynamic power spectrum of the XMM-Newton observation of 4U 1636--53. The gray vertical line indicates the time of an X-ray burst in the observation. Each column represents the power spectrum extracted from a time interval of 1130 seconds with the starting time of each segment set to 565 s after the starting time of the previous segment. The frequency scale was oversampled by a factor of 100 using the Lomb-Scargle periodogram to display the evolution of the mHz QPO frequency. The count rate in bad time intervals due to the X-ray burst and instrument dropouts was fixed to the average count rate of 340 counts s$^{-1}$. In total, there are 71 power spectra shown in the plot. The colour bar on the right indicates the power at each frequency as defined in the Lomb-Scargle periodogram.}
\label{dps}
\end{figure}

To account for the possible evolution of the frequency of this QPO, which we call the mHz QPO, we divided the observation into 71 overlaping time intervals and calculated a dynamic power spectrum oversampling the frequency scale by a factor of 100 using the Lomb-Scargle periodogram method \citep{lomb76,scargle82}. Each time interval in the dynamic power spectrum was 1130 seconds long, with the next interval starting 565 s after the start time of the previous interval. We set the count rate of bad time intervals due to the X-ray burst and instrument dropouts in the light curve to the average count rate (340 counts s$^{-1}$), the final dynamic power spectrum is shown in Figure \ref{dps}. 

From Figure \ref{dps} it is apparent that the mHz QPO is present only at the beginning and at the end of the observation. Guided by this plot, we divided the whole observation into four segments, three of them (S1$-$S3) of similar length ($\sim$12 ks), and the last one (S4) covering the remaining $\sim$ 3.5 ks of the observation. The QPO is present in the first and the last segments (S1 and S4), while it is not detected in the second and third ones (S2 and S3; see \S 3.1). We extracted a light curve and an average power spectrum from each segment; the details of the four segments are listed in Table \ref{overall}. 

We further divided the first segment, before the X-ray burst, into three subsegments, S1a, S1b and S1c, according to the behaviour of the QPO frequency, and generated one light curve for each subsegment. The time intervals of each subsegment were S1a: $0-3954$ s, S1b: $3954-7344$ s and S1c: $7344-11579$ s, respectively, from the start of the observation (MJD=55079.05539). 

We fitted the total average power spectra using the model {\sc constant+lorentz+lorentz}. The {\sc constant} describes the Poisson noise level, and the first {\sc lorentz} component represents the QPO around 7$-$8 mHz, while the second one represents its second harmonic component. We linked the frequency and width of the second {\sc lorentz} component to be the double and the same as in the first {\sc lorentz} component, respectively. For segments 2 and 3, where no QPO is detected, we fitted the average power spectra using the model {\sc constant+lorentz} to calculate the upper limit of the rms amplitude of the QPO. We fixed the frequency and the width of the {\sc lorentz} component to be the values in the total observation since they are poorly constrained due to the short time ranges ($<$12 ks). For the three subsegments and the last segment, due to their limited observational time span ($\le$4.5 ks), we explored the timing properties of the QPO from the light curves: We fitted the light curve of each independent 1130 s time interval in S1 and S4 using a model consisting of a constant plus a sine function. And then we calculated the average frequency and its standard deviation in S1a, S1b, S1c and S4 from the frequencies in the corresponding intervals. Furthermore, we used average frequency to generate a folded light curve for each subsegments and the last segment, and derived the rms amplitude from the fits of the folded light curves.  

\section{Results}
\subsection{Timing results}
In Figure \ref{lc} we plot the light curve of the observation. The length of the light curve is about 40.8 ks, with an average count rate of about 340 counts s$^{-1}$. In Figure \ref{ps} we show the average power spectrum of the total observation. We found a strong QPO at around 7$-$8 mHz and a possible second harmonic component at $\sim 15$ mHz. In the dynamic power spectrum we detect the mHz QPO and the second harmonic from the start of the observation until an X-ray burst occurs (around 12 ks in the plot). After the burst, the mHz QPO disappears for about 25.3 ks, and then reappears in the last 3.5 ks of the observation. We estimated the significance of the signal against the hypothesis of white noise via the Lomb-Scargle method. The false alarm probability of the QPO detection (without considering the harmonic) in the total observation is 2.21$\times 10^{-8}$, while the probability is 5.24$\times 10^{-19}$ for segments 1 and 4. 

For the total observation, the central frequency of the {\sc lorentz} component is 7.2$_{-0.2}^{+0.4}$ mHz, with a full-width at half maximum of 2.3$_{-0.7}^{+1}$ mHz. The rms amplitude is 1.11$\pm 0.31$\%, which is consistent with the previous measurements \citep{revni01,diego08}, although note that \citet{revni01} and \citet{diego08} used RXTE data which samples $>$ 3 keV range, while we use 0.8$-$11 keV here. In segments 2 and 3, where no QPO is significantly detected, the 95\% upper limit of the rms amplitude is 0.49\% and 0.40\%, respectively. The average frequency of the mHz QPO decreases from 8.3 mHz in S1a to 6.7 mHz in S1b, and increases slightly to 6.9 mHz in the S1c, before the X-ray burst. After the burst, the QPO reappears at an average frequency of 9.4 mHz in segment 4. The rms amplitude of the QPO increases from 0.80$\pm 0.10$\% in S1a to 2.27$\pm 0.10$\% in S1c, and then it is 1.34$\pm 0.10$\% in segment 4, when it reappears $\sim$ 25 ks after the X-ray burst. The parameters of the mHz QPO are given in Table \ref{qpo}.

\begin{table}
\caption{Parameters of the QPO in the first three subsegments (S1a, S1b and S1c) and the last segment of the observation of 4U 1636$-$53. Here we give the standard deviation of the average frequency due to its large variation in the subsegments.  All errors and upper limits in the Tables are, respectively, at the 90\% and 95\% confidence level unless otherwise indicated.}
\begin{tabular}{|c|c|c|}
\hline
\hline
  &  Average frequency (mHz)    &       RMS amplitude (\%)    \\
\hline
S1a &  8.3$\pm 0.6$&  0.80$\pm 0.10$    \\   
S1b &  6.7$\pm 0.2$&  1.48$\pm 0.12$   \\
S1c &  6.9$\pm 0.2$&  2.27$\pm 0.10$    \\
S4  &  9.4$\pm 0.3$&  1.34$\pm 0.10$   \\
\hline
\end{tabular}
\medskip
\label{qpo}
\end{table}

\subsection{Spectral results} 
  In all our fits, we first tried either a single thermal component (blackbody or disc), or a single Comptonised component to describe the continuum; the results, however, showed that the spectra could not be well fitted with only one of those component. We found that both a thermal and a Comptonised component were necessary to fit the continuum, whereas two thermal components plus a Comptonised component \citep[e.g.,][]{sanna13,lyu14} was too complex a model given the quality of the data. For the thermal component either a blackbody or an accretion-disc model could fit the data well. We used either the {\sc bbody} component in XSPEC 12.8.1 \citep{arnaud96} to describe the combined thermal emission from the neutron star surface and its boundary layer, or the {\sc diskbb} component \citep{mitsuda84,maki86} to model the multi-temperature thermal emission from an accretion disc. The Comptonised component was modelled by {\sc nthcomp} \citep{zdzi96,zyck99}, which describes the emission originating from the inverse Compton scattering process in the corona, with the seed thermal photons coming either from the accretion disc or the neutron star surface (plus boundary layer). In this work we chose the disc as the source of seed photons for the scattering in the corona \citep{sanna13,lyu14}. We also included the component {\sc phabs} in all our models to account for absorption from the interstellar material along the line of sight. For this component we used the solar abundance table of \citet{wilms00} and the photoionization cross section table of \citet{verner96}. We added a 0.5 \% systematic error to the model.   
  
   After fitting the spectra with these continuum models in all cases we found significant residuals around $6-7$ keV, where possible emission from an iron line is expected \citep{pandel08,cackett10,sanna13,lyu14}, we therefore added a line component to the model with the central energy of the line constrained to the range 6.4 $-$ 6.97 keV. We first used a Gaussian line to estimate some general properties of the iron line. The fitted width of the line ($\sigma$) was between 1.1 keV and 1.6 keV, indicating that a broadening mechanism was required to explain the line profile. We therefore also used the emission line model, {\sc kyrline} \citep{dovciak04}, instead of the Gaussian to fit the iron line. The {\sc kyrline} component describes a relativistic line from an accretion disc around a black hole with arbitrary spin. In this work we fixed the spin parameter at 0.27, which was derived from the spin frequency of the neutron star (see \citet{sanna13} and \citet{lyu14} for details), while the outer radius of the disc was fixed at $1000 GM/c^{2}$, where G is the gravitational constant, M is the mass of neutron star and c is the speed of light.

We noticed that the column density, $N_{H}$, in {\sc phabs} and the
power-law index, $\Gamma$, the temperature of the disc seed photons,
$kT_{dbb}$, and the electron temperature, $kT_{e}$, in {\sc nthcomp}
changed systematically between the fits with the two iron line models,
({\sc gauss} or {\sc kyrline}), which in turn changed the normalisation
of the spectral components from one model to the other. These
differences, however, did not alter the general trends of the parameters
as a function of segment number, but just shifted the relations
consistently up or down. These variations are likely due to the lack of
data above 12 keV, which leads to some degeneracy between parameters
(especially the power-law index and electron temperature of the corona).
To mitigate this problem we proceeded as follows: 

We defined four models, A to D, in Xspec; the continuum components were {\sc phabs}, {\sc bbody} and {\sc nthcomp}
for models A and B, and {\sc phabs}, {\sc diskbb} and {\sc nthcomp} for models C and D, respectively. For model A and C,
we fitted the line with {\sc gauss}, and for B and D with {\sc kyrline}. For models A and B we fitted the spectra of all 
segments simultaneously linking $N_{H}$, $\Gamma$, $kT_{dbb}$ and $kT_{e}$ across segments in each model separately. For models C
and D we did the same, except that we did not link $kT_{dbb}$. For models B and D, for which we used {\sc kyrline}, we further linked the inclination angle of
the accretion disc, $\theta$, to be the same in all segments. The other parameters (see Table \ref{unlink1} and Table
\ref{unlink2}) were left free between models and segments. After we found the best-fitting parameters for the joint fits with models A and
B, we deleted model B (model A) and verified that the parameters obtained from the joint fits gave acceptable fits for model A (model B)
only. We did the same for models C and D.

\begin{table*}
\caption{Fitting results of 4U 1636$-$53, linked parameters. Values and errors in the Tables are round off to the first or the second decimals.}
\begin{tabular}{|c|c|c|c|c|}
\hline
\hline                       
\multirow{2}{*}{Model comp}&\multirow{2}{*}{Parameter}& \multicolumn{2}{|c|}{Four segments}&\multicolumn{1}{|c|}{Three subsegements}\\
 &    & {\sc bbody + nthcomp}$^{*}$ & {\sc diskbb + nthcomp}$^{*}$ & {\sc bbody + nthcomp}\\
\hline                                       
{\sc phabs  } &   $N_{H} (10^{22}) $&  0.38$_{-0.02}^{+0.01}   $&  0.27$ \pm 0.01       $&0.392$_{-0.03}^{+0.004}  $\\                 
{\sc kyrline} &   $\theta$ (deg)    &  86.2$\pm 0.9            $&  85.9$ \pm 1.1        $&86.5$_{-0.2}^{+1.0}     $\\                 
{\sc nthcomp} &   $\Gamma          $&  1.88$\pm 0.01           $&  2.16$_{-0.07}^{+0.02}$&1.92$\pm 0.01             $\\                
              &   $kT_{e} (keV)    $&  3.08$\pm 0.08           $&  5.5$_{-1.1}^{+0.6}  $&3.29$_{-0.01}^{+0.2}    $\\                 
              &   $kT_{dbb} (keV)  $&  0.24$_{-0.02}^{+0.07}   $&  -$^{ \flat}$                      &0.22$_{-0.01}^{+0.03}    $\\      
              &$\chi^2_\nu$ $(\chi^2/dof)$&     1.09 (1463/1347)                  &     1.16 (1568/1348)                    &     1.01 (1017/1002)                     \\                 
\hline             
\hline             
\end{tabular}      
\medskip           
\\
$\ast$ We used two model combinations, {\sc nthcomp + bbody} and {\sc nthcomp + diskbb}, to fit the continuum of the four segments. \\
$\flat$ The temperature of the {\sc diskbb} component, not linked in the fit, is therefore shown in Table \ref{unlink2}. 
\label{link1}    
\end{table*}

  To summarise: We used two models, {\sc nthcomp + bbody} or {\sc nthcomp + diskbb}, to fit the continuum, and either a {\sc gauss} or {\sc kyrline} component to describe the iron emission line for the spectra of the four segments. Besides, we fitted all spectra with the two line models simultaneously, linking some parameters between the models and the segments. 

\begin{figure}
\center
\includegraphics[height=0.38\textwidth]{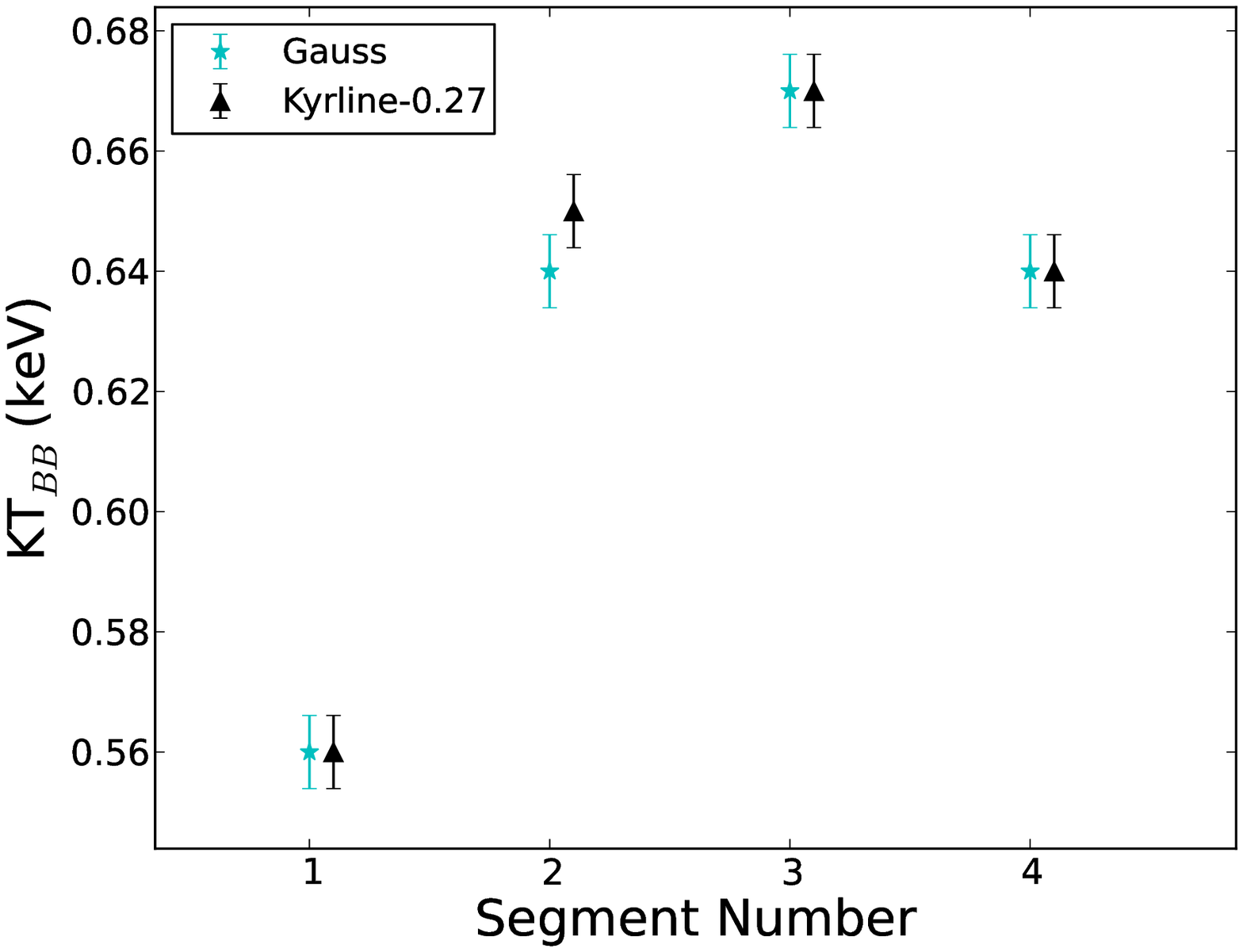}
\caption{Temperature of the blackbody component in 4U 1636$-$53 as a function of segment number for the XMM-Newton observation. The model {\sc bbody+nthcomp} was used to fit the continuum. Different colours/symbols show the results of fits with different models to the iron line, as indicated in the legend, with Kyrline-0.27 representing the results when we fitted the line with a {\sc kyrline} model with the spin parameter fixed to 0.27. In this and other plots an offset in the $x$-axis between the results of the two iron line models has been added for clarity. Error bars in this and the other plots correspond to the 1-$\sigma$ confidence range.}
\label{par1}
\end{figure}

  Here we only describe the main results of the fit with the {\sc bbody} component; The results of the fit with the {\sc diskbb} component are shown in Appendix.

\begin{figure}
\center
\includegraphics[height=0.5\textwidth]{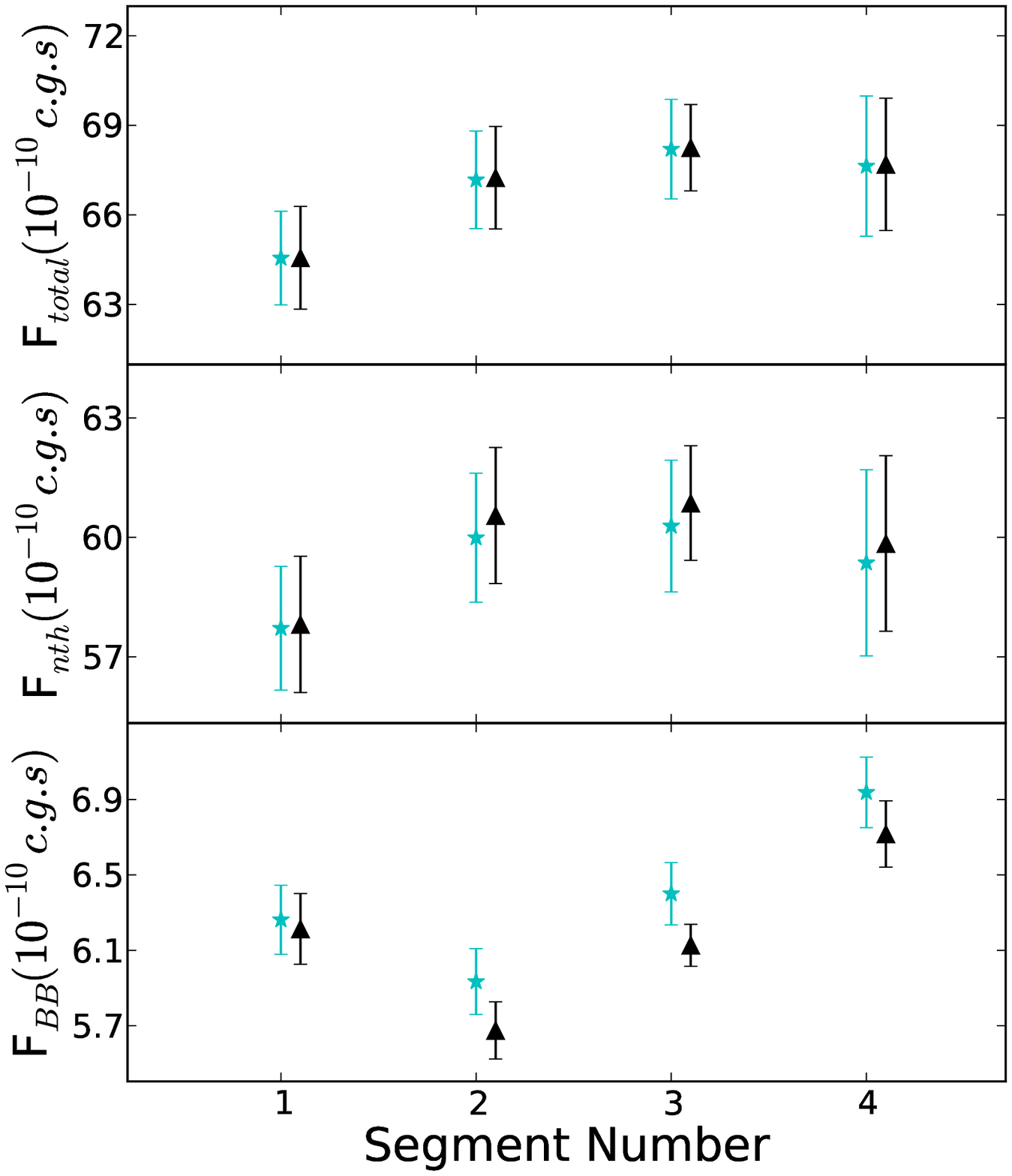}
\caption{The total unabsorbed flux (top panel), the flux of the {\sc nthcomp} component (middle panel) and the {\sc bbody} component (bottom panel) in the $0.5-130$ keV range for 4U 1636$-$53 as a function of segment number. The model {\sc bbody+nthcomp} was used to fit the continuum. Symbols are the same as in Figure \ref{par1}.}
\label{flux1}
\end{figure}

\begin{table*}
\small
\caption{Fitting results of the four segments of 4U 1636$-$53 using the continuum model {\sc bbody+nthcomp}, unlinked parameters. We fit the four segments with two iron line models simultaneously (see text for more details); the $\chi^2_\nu$ ($\chi^2/dof)$ of the fit is 1.09 (1463/1347). }
\begin{tabular}{|c|c|c|c|c|c|}
\hline
\hline
Model comp    & Parameter        &   S1                   &S2                 &S3                   &S4                      \\
\hline
{\sc bbody}  &  $kT$ (keV)                &  0.56$\pm 0.01     $& 0.65$\pm 0.01   $& 0.67$\pm 0.01  $ & 0.64$\pm 0.01   $  \\
             &  Normalization (10$^{-3}$)  &  7.2$\pm 0.4       $& 6.6$\pm 0.3    $& 7.1$\pm 0.2    $ & 7.8$\pm 0.3    $  \\
             &  Flux (10$^{-10}$)          &6.2$\pm 0.3         $&  5.7$\pm 0.2   $& 6.1$\pm 0.2    $ & 6.7$\pm 0.3$ \\
{\sc nthcomp}&  Normalization            &  0.87$\pm 0.03      $& 0.91$\pm 0.03   $& 0.91$\pm 0.02  $ & 0.90$\pm 0.05   $  \\
             &  Flux (10$^{-10}$)          &57.8$\pm 2.8       $&  60.5$\pm 2.8   $& 60.9$\pm 2.4   $&  59.8$\pm 3.6 $ \\
{\sc kyrline}&$R_{\rm in}$ ($R_{G}$)&  11.5$\pm 2.3       $& 5.12$_{-0}^{+0.27}$& 5.12$_{-0}^{+0.12}$ & 5.12$_{-0}^{+0.41}$  \\
             &$E_{\rm line}$ (keV)&  6.97$_{-0.15}^{+0}  $& 6.59$\pm 0.14   $& 6.60$\pm 0.09   $ & 6.85$\pm 0.12   $  \\
             &  $\alpha$        &  3.2$\pm 0.5       $& 2.8$\pm 0.2    $& 2.8$\pm 0.2    $ & 2.6$\pm 0.2    $  \\
             &  Normalization (10$^{-3}$)  &  4.7$\pm 0.6       $& 9.3$\pm 0.7    $& 11.5$\pm 0.8    $ & 9.9$\pm 1.2    $   \\
             &  Flux (10$^{-10}$)          &0.53$\pm 0.07 $& 1.02$\pm 0.09$& 1.26$\pm 0.09 $&  1.1$\pm 0.1$ \\
             &  Total flux (10$^{-10}$)    & 64.6$\pm 2.8 $&  67.2$\pm 2.8 $&  68.3 $\pm 2.4 $&  67.7$\pm 3.6 $\\                 
\hline                                             
{\sc bbody}  &  $kT$ (keV)        & 0.56$ \pm 0.01  $&  0.64$ \pm 0.01  $& 0.67$ \pm 0.01 $&  0.64$\pm 0.01    $  \\
             &  Normalization (10$^{-3}$)  & 7.3$ \pm 0.4    $&  6.9$ \pm 0.3   $& 7.4$ \pm 0.3   $&  8.0$\pm 0.4     $  \\
             &  Flux (10$^{-10}$)          &6.3$\pm 0.3$& 5.9$\pm 0.3$& 6.4$\pm 0.3$& 6.9$\pm 0.3$ \\
{\sc nthcomp}&  Normalization   & 0.87$ \pm 0.03   $&  0.90$ \pm 0.03  $& 0.91$ \pm 0.03  $&  0.89$\pm 0.05    $  \\
             &  Flux (10$^{-10}$)          &57.7$\pm 2.6 $&  60.0$\pm 2.7$&  60.3$\pm 2.7 $& 59.4$\pm 3.8$ \\
{\sc gauss} &$E_{\rm line}$ (keV)& 6.97$_{-0.08}^{+0}$&  6.78$ \pm 0.17  $& 6.84$ \pm 0.13  $&  6.97$_{-0.12}^{+0}$  \\
             &  $\sigma$ (keV)  & 1.1$ \pm 0.1    $&  1.6$ \pm 0.2   $& 1.5$ \pm 0.1   $&  1.5$\pm 0.2     $  \\
             &  Normalization (10$^{-3}$)  & 5.1$ \pm 0.8    $&  11.3$ \pm 1.9  $& 13.7$ \pm 1.8   $&  11.8$\pm 1.9     $  \\
             &  Flux (10$^{-10}$)          &0.57$\pm 0.08 $&  1.2$\pm 0.2$ &  1.5$\pm 0.2$ &  1.3$\pm 0.2$ \\
             &  Total flux (10$^{-10}$)    &64.6$\pm 2.6 $&  67.2$\pm 2.7$ &  68.2$\pm 2.7$ &  67.6$\pm 3.9 $\\
\hline
\hline
\end{tabular}
\medskip
\label{unlink1}
\end{table*}

   In Figure \ref{par1} we plot the evolution of the temperature of the {\sc bbody} component for the four segments. The blackbody temperature, $kT_{bb}$, first increases from 0.56$\pm$0.01 keV to 0.67$\pm$0.01 keV from segment 1 to segment 3, and then decreases to 0.64$\pm$0.01 keV in the last segment (all errors represent the 90\% confidence range). It appears that $kT_{bb}$ increases from the QPO segment to the non-QPO segment (from S1 to S2), and vice versa (from S4 to S3).
     
  In Figure \ref{flux1} we show the total unabsorbed flux ($0.5-130$ keV), the unabsorbed flux of the {\sc nthcomp} component and that of the {\sc bbody} component as a function of the segment number. The total flux increases from segment 1 to segment 2, and remains more or less constant in the last three segments, while the flux of {\sc nthcomp} increases from the first segment to the second one, and then remains constant or decreases marginally. The flux of the {\sc bbody} component decreases from segment 1 to segment 2, and then increases in the last three segments.   

  The best fitting parameters to the spectra of the four segments using {\sc bbody} to fit the soft component are shown in Table \ref{link1} (linked parameters: $N_{H}$, $\Gamma$, $kT_{e}$, $\theta$ and $kT_{dbb}$) and Table \ref{unlink1} (unlinked parameters), respectively. All errors and upper limits in the Tables are, respectively, at the 90\% and 95\% confidence level, unless otherwise indicated. 
   
As mentioned in \S 2.2, we divided the first segment into three subsegments, and we analysed the spectra of S1a, S1b and S1c in the same way as we did for the four segments, except that here we only used the {\sc nthcomp + bbody} model to describe the continuum, since the mHz QPOs are regarded to be connected to marginal stable burning on the neutron star surface (see \S 1). The best fitting parameters to the spectra of the three subsegments are given in Table \ref{link1} (linked parameters: $N_{H}$, $\Gamma$, $kT_{e}$, $\theta$ and $kT_{dbb}$) and Table \ref{unlink3} (unlinked parameters), respectively.

\begin{table*}
\small
\caption{Fitting results of the first three subsegments of 4U 1636$-$53 using the continuum model {\sc bbody+nthcomp}, unlinked parameters. We fit the three subsegments with two iron line models simultaneously (see text for more details); the $\chi^2_\nu$ $(\chi^2/dof)$ of the fit is 1.01 (1017/1002). }
\begin{tabular}{|c|c|c|c|c|}
\hline
\hline
Model comp    & Parameter        &   S1a                   &S1b                 &S1c                                  \\
\hline
{\sc bbody}  &  $kT$ (keV)        & 0.60$\pm 0.02 $&  0.57$\pm 0.02 $&  0.59$\pm 0.02  $                                                                                 \\
             &  Normalization (10$^{-3}$)  & 7.7$\pm 0.4          $&  7.2$\pm 0.2          $&  7.18$_{-0.06}^{+ 0.5}   $    \\
             &  Flux (10$^{-10}$)          & 6.7$\pm  0.3  $&    6.2$\pm 0.2  $&   6.2$\pm  0.2  $    \\             
{\sc nthcomp}&  Normalization   & 0.91$\pm 0.04         $&  0.90$\pm 0.03         $&  0.91$_{-0.06}^{+0.01}  $                                                                                  \\
             &  Flux (10$^{-10}$)          & 58.3 $\pm   2.4 $&     58.1 $\pm  2.5 $&    58.4$\pm  2.4$   \\
{\sc kyrline}&  $R_{\rm in}$ ($R_{G}$)& 5.12$_{-0}^{+1.28}     $&  5.12$_{-0}^{+0.52}     $&  10.0$_{-1.6}^{+0.3}$                                                                                \\
             &  $E_{\rm line}$ (keV)& 6.46$\pm 0.08         $&  6.5$\pm 0.1           $&  6.97$_{-0.24}^{ +0 }     $                                                                              \\
             &  $\alpha$        & 2.9$\pm 0.2          $&  2.5$\pm 0.2           $&  3.9$\pm 0.3            $                                                                                \\
             &  Normalization (10$^{-3}$)  & 10.23$_{-0.02}^{+0.8} $&  7.3$_{-0.3}^{+1.2}    $&  7.5$\pm 1.0            $         \\                                                                       
             &  Flux (10$^{-10}$)          &1.09$\pm  0.05 $&     0.78 $\pm 0.08 $&    0.8 $\pm 0.1 $   \\
             &  Total flux (10$^{-10}$)    &66.1$\pm  2.4 $&     65.0 $\pm 2.5 $&    65.4 $\pm 2.4  $\\           
\hline                                                                                                                                     
{\sc bbody}  &  $kT$ (keV)        & 0.60$\pm 0.01 $&  0.58$\pm 0.01          $&  0.59$\pm 0.01  $                                                                                 \\
             &  Normalization (10$^{-3}$)  & 7.8$\pm 0.5 $&  7.2$\pm 0.3           $&  7.4$\pm  0.2           $                                                                                \\
             &  Flux (10$^{-10}$)          &6.7 $\pm 0.4  $&    6.2$\pm  0.2 $&    6.4 $\pm 0.2 $ \\
{\sc nthcomp}&  Normalization   & 0.90$_{-0.06}^{+0.01}  $&  0.90$_{-0.06}^{+0.01} $&  0.90$\pm  0.03          $  \\   
             &  Flux (10$^{-10}$)          &58.2 $\pm 2.5  $&    58.1 $\pm 2.5  $&   58.0 $\pm  2.3  $ \\
{\sc gauss} &$E_{\rm line}$ (keV)& 6.77$_{-0.04}^{+0.07}  $&  6.79$_{-0.1}^{+0.07}   $&  6.88$_{-0.2}^{+0.09}    $                                                                                 \\
             &  $\sigma$ (keV)  & 1.37$_{-0.03}^{+0.1}   $&  1.19$_{-0.2}^{+0.07}   $&  1.34$_{-0.01}^{+0.09}   $                                                                                 \\
             &  Normalization (10$^{-3}$)  & 10.5$_{-0.6}^{+1.4}    $&  7.1$_{-0.1}^{+1.5}    $&  8.94$_{-0.3}^{+ 1.5}    $                                                                                \\
             &  Flux (10$^{-10}$)          &1.2 $\pm 0.1  $&    0.78 $\pm 0.09 $&    1.0 $\pm  0.1 $ \\
             &  Total flux (10$^{-10}$)    &66.1 $\pm 2.6   $&   65.1 $\pm 2.5 $&    65.4 $\pm 2.3 $ \\
\hline
\hline
\end{tabular}
\medskip
\label{unlink3}
\end{table*}

\begin{figure}
\center
\includegraphics[height=0.45\textwidth]{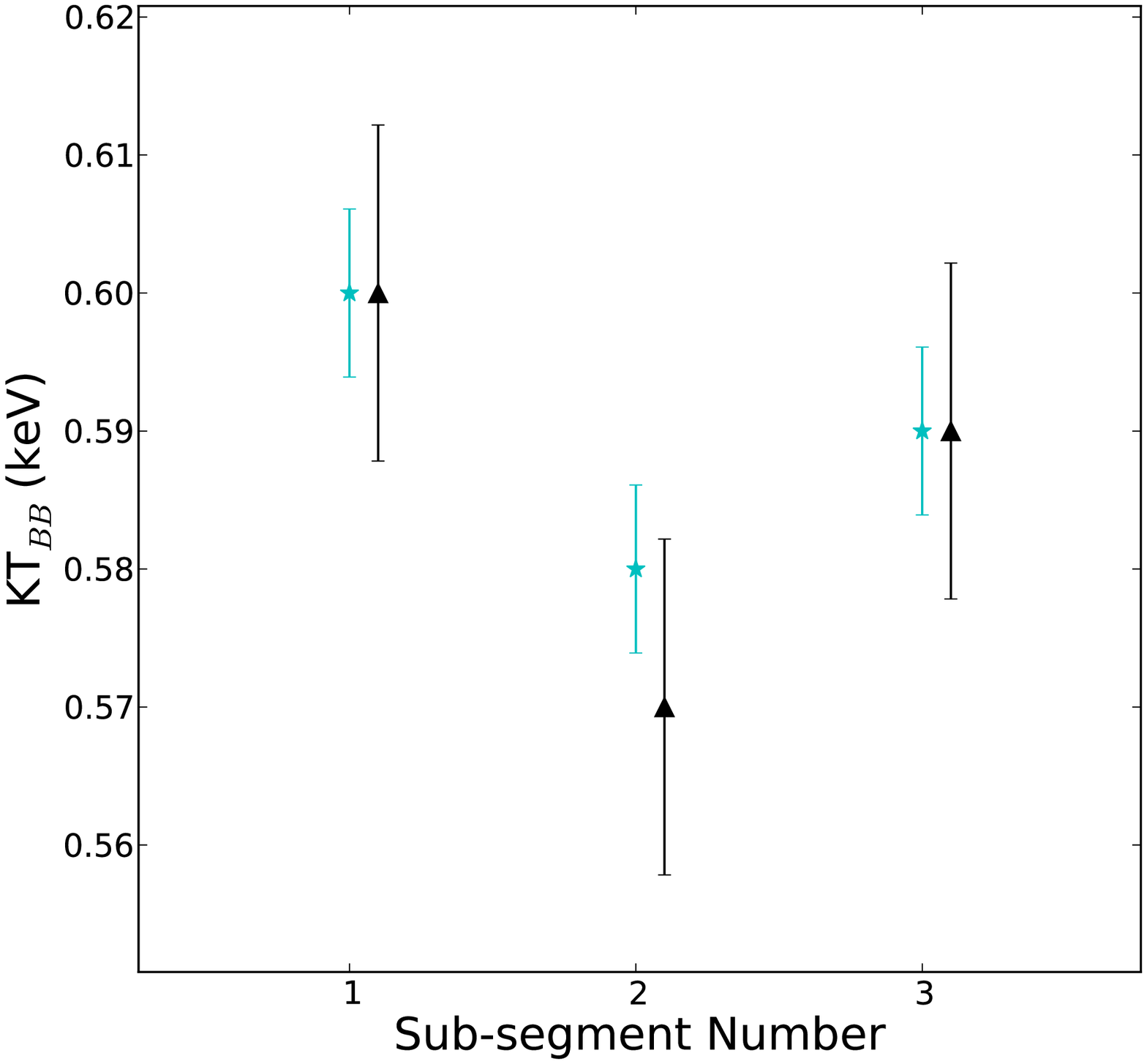}
\caption{Temperature of the blackbody component in 4U 1636$-$53 for the first three subsegments (S1a, S1b and S1c) of the XMM-Newton observation. The model {\sc bbody+nthcomp} was used to fit the continuum. Symbols are the same as in Figure \ref{par1}.}
\label{s1}
\end{figure}

In Figure \ref{s1} we plot the blackbody temperature, $kT_{bb}$, vs. subsegment number. The blackbody temperature is 0.60$\pm 0.01$ keV for the {\sc gauss} (0.60$\pm 0.02$ keV for the {\sc kyrline}) in the S1a, it decreases to 0.58$\pm 0.01$ keV (0.57$\pm 0.02$ keV for the {\sc kyrline}) in S1b, and finally it changes to 0.59$\pm 0.01$ keV (0.59$\pm 0.02$ keV for the {\sc kyrline}) in the S1c.

\begin{figure}
\center
\includegraphics[height=0.5\textwidth]{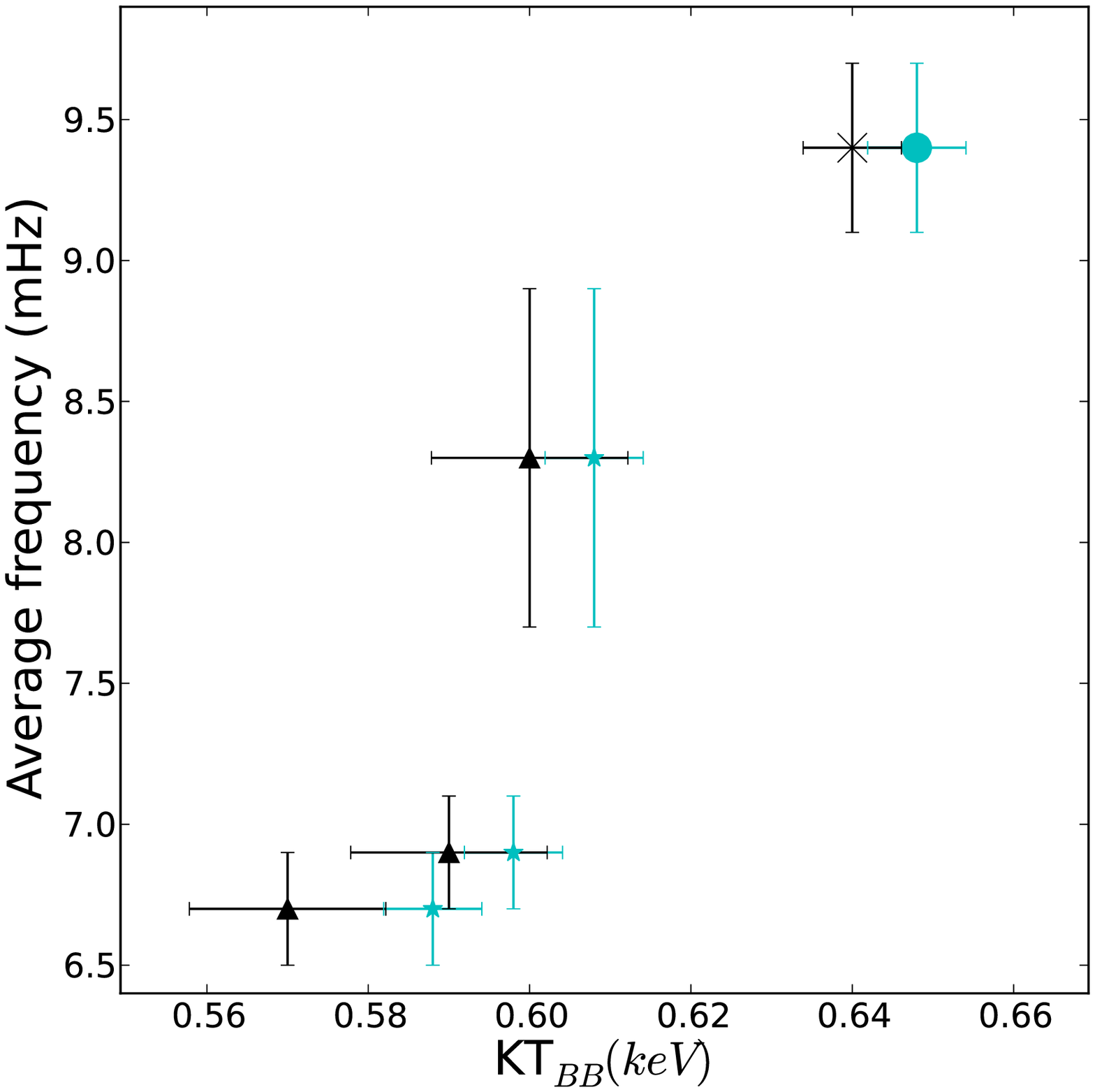}
\caption{Average frequency of the mHz QPO vs. the temperature of the blackbody component for the first three subsegments and the last segment of the observation of 4U 1636$-$53. The model {\sc bbody+nthcomp} was used to fit the continuum. Here we plot the standard deviation of the average frequency. We added an offset of 0.008 keV to the temperature of the blackbody component in the case of {\sc gauss} model for clarity. Symbols are the same as in Figure \ref{par1}. We use an "X" symbol in black and a filled circle in cyan to show the results of the {\sc kyrline} and the {\sc gauss} iron line model for the segment 4, respectively.}
\label{sc1}
\end{figure}         

In Figure \ref{sc1} we show the relation between the average frequency of the mHz QPO and the temperature of the blackbody component in the three subsegments and the last segment, where the QPO reappears after the burst. Considering the variation of the frequency in the first subsegment, we plot the standard deviation of the average frequency as error bars in the plot. This Figure shows a clear correlation between the average frequency of the QPO and the temperature of the blackbody component, the correlation coefficient between the frequency and the temperature is 0.95.

\section{discussion}
We report the first detection of a mHz quasi-periodic oscillation (QPO), in the neutron-star LMXB 4U 1636--53, with XMM-Newton. At the beginning of the observation the frequency of the mHz QPO was $\sim$$8.3$ mHz, and then the frequency slowly decreased to below $\sim$7 mHz. At the time of $\sim$12 ks a PRE X-ray burst occurs \citep{zhang11}, the mHz QPO disappears, and it subsequently reappears about 25.3 ks later at a frequency of $\sim$9.4 mHz. This is the longest time interval so far measured between the disappearance and reappearance of mHz QPO in any LMXB. The $0.8-11$ keV rms amplitude of the QPO increases from 0.80$\pm 0.06$\% in S1a to 2.27$\pm 0.06$\% in S1c just before the X-ray burst, and changes to 1.34$\pm 0.06$\% when the QPO reappears after the burst. Finally, we discovered a strong correlation between the frequency of the mHz QPO and the temperature of the neutron-star surface, represented by a blackbody component in the energy spectrum of the source. 

Using RXTE observations, \citet{diego08} found that the mHz QPOs in 4U 1636--53 occur only when the source is in certain spectral states: When the source is in an intermediate spectral state, close to the transition between the soft and the hard state \citep[see Figure 1 in][]{diego08}, 4U 1636--53 exhibits mHz QPOs with frequencies that systematically decrease with time from $\sim$15 mHz to around $7-9$ mHz, and at that frequency the QPOs disappear simultaneously with the occurrence of a Type I X-ray burst. When 4U 1636--53 is in a softer state, the oscillations show no frequency drift, and the frequency is always constrained between $\sim$7 and $\sim$9 mHz.

The QPO we found in the XMM-Newton observation of 4U 1636--53 shows a frequency drift, from $\sim$8.3 mHz to $\sim$7 mHz, and disappears after the Type-I X-ray burst onset. As previously reported by \citet{sanna13} and \citet{lyu14}, the XMM-Newton observation analysed in this paper \citep[Obs.5 in][]{sanna13} sampled the transition between the soft state and the hard state of 4U 1636--53 \citep[see Figure 2 in][]{sanna13}. Our results are therefore consistent with the picture that frequency drifts of the mHz QPOs are only observed in the state transition.

\citet{diego08} also showed that the time interval required to detect the mHz QPOs after an X-ray burst is not always the same, and it is independent of the spectral state of the source. However, due to gaps in the RXTE observations, they could not exclude waiting times as short as $\sim$1000 s. \citet{molkov05} observed a $\sim$10 mHz oscillations in the LMXB SLX 1735--269 that reappeared during the decaying phase of an X-ray burst, just $\sim$400 s after the peak of the burst. At the other extreme, \citet{diego08} found that no mHz QPOs were detected during the $\approx$15 ks of uninterrupted data after an X-ray burst in 4U 1636--53, setting a lower limit on the longest waiting time for this source. In this paper, and thanks to the ability of XMM-Newton to carrying out long observations without data-gaps, we find a waiting time of $\sim$25 ks between an X-ray burst and the reappearance of the mHz QPO, the longest waiting time so far.

Standard theory predicts that the transition from stable to unstable helium burning via the triple-alpha process takes place when the accretion rate is close to the Eddington limit \citep{fujimoto81,ayasli82}. If the accreted matter is hydrogen-deficient, the transition is expected to take place at an even higher accretion rate \citep{bildsten98,keek09}. 4U 1636--53 accretes at few percent of Eddington; however if local mass accretion rate per unit area is $\dot{m}\simeq\dot{m}_{Edd}$, as suggested by \citet{heger07}, only $\approx1000$ seconds are required to accrete a fuel layer of column depth $y_f$ capable of undergoing marginally stable burning \citep[assuming that none of the accreting fuel is burnt, $y_f\approx10^8$~g~cm$^{-2}$, and $\dot{m}\approx$ 8$\times 10^4$~g~cm$^{-2}$~s$^{-1}$; see, e.g.][]{heger07}. Waiting times longer than 1000 s, and particularly as long as $\sim$25 ks, imply that at the observed luminosities either a large fraction of the accreted fuel must be burnt as it accretes onto the neutron-star surface \citep[e.g.,][and references therein]{ vanparadijs88,cornelisse03,galloway08, diego08, keek09}, or that only a small fraction of the accreted matter reaches the neutron-star surface.

If hydrogen and/or helium can mix efficiently as mass is accreted onto the neutron star, the conditions under which the fuel is burnt are different \citep[e.g.,][]{fujimoto93,yoon04,piro07,keek09}. \citet{keek09} studied the effect of rotationally induced transport processes on the stability of helium burning. They found that as helium is diffused to greater depths, the stability of the burning is increased, such that the critical accretion rate for stable helium burning decreases and, combined with a higher heat flux from the crust, turbulent mixing could explain the fact that mHz QPOs occur at (apparent) lower accretion rates. Furthermore, \citet{keek09} found that by lowering the heat flux from the crust (effectively cooling the burning layer), the frequency of the marginally-stable burning decreases. 4U 1636--53 shows mHz QPOs with systematically decreasing frequencies right before a type I X-ray burst \citep{diego08}. That behaviour of the frequency is considered to be related to the heating of the deeper layers of the neutron star. Here we find a strong correlation between the average frequency of the mHz QPO and the temperature of the blackbody component in the energy spectrum. It is worth mentioning that the blackbody temperature from the fits corresponds to the effective temperature of the neutron-star photosphere rather than the temperature of the burning layer on the neutron star surface. Besides, the blackbody temperature could in fact be partly the temperature of the disc, since we do not fit a disc component and a blackbody component separately, but just a blackbody component that accounts for both components. If the temperature of the burning layer and that of the photosphere are correlated, and the temperature of the disc remains more or less constant during our observation, the frequency of the mHz QPO should be correlated with the temperature of the burning layer. This is different from the prediction of the model of \cite{heger07}, in which the frequency is inversely proportional to the square root of the temperature of the burning layer. Assuming that mass accretion rate is proportional to the total flux in each segment (see Fig \ref{flux1}), and the thickness of the fuel layer is more or less constant, we found no significant correlation between the observed frequency and the frequency predicted by eq. (11) in \cite{heger07}. In our data the predicted frequency changes by $\sim$0.5\%, but the errors are about 5\%, which precludes us from drawing any conclusion. On the other hand, a correlation between QPO frequency and neutron-star temperature is consistent with the results of \citet{keek09}: When the burning layer is effectively cooled down, the frequency of the oscillations decreases by tens of percents until a flash occurs. Interestingly, we find that the frequency-temperature correlation holds also when the mHz QPO reappears after the burst. Additionally, the model of \citet{keek09} could also explain why the mHz QPOs are not sensitive to short-term variations in the accretion rate: The cooling could, for instance, be dominated by the slow release of energy from a deeper layer that was heated up during an X-ray burst \citep{keek09}.

Furthermore, \citet{keek09} found that the flux from the neutron-star crust, $L_{critical}$, at the transition between stable and unstable burning increases with increasing accretion rate. This may offer a clue about the mechanism that triggers the reappearance of the mHz QPO after the burst. As shown in Figure \ref{flux1}, the total flux in segments 2 to 4, after the X-ray burst, is higher than the total flux in the first segment, before the burst, indicating that the accretion rate may have increased after the burst. If $L_{critical}$ in segment 1 is $L^{lo}_{critical}$, the increase of accretion rate after the X-ray burst would lead to a higher value, $L^{hi}_{critical}$, which in turn sets a higher threshold that needs to be overcome for the mHz QPO to reappear. The flux of the blackbody component (lower panel in Figure \ref{flux1}) in segment 1 is around 6.2$\times$10$^{-10}$ ergs cm$^{-2}$ s$^{-1}$; after the burst the flux drops in segment 2, and after that it increases until the end of the observation. Only segment 4 shows a blackbody flux that is significantly larger than the one in segment 1, indicating that the heat flux from the crust in segment 4 may be as high as the value of $L^{hi}_{critical}$ required for the mHz QPO to reappear. This scenario is also consistent with the prediction of the standard theory: After an X-ray burst fuel on the neutron-star surface becomes hydrogen-deficient, leading to a transition at a higher accretion rate.

\section*{Acknowledgments}

This research has made use of data obtained from the High Energy Astrophysics Science Archive Research Center (HEASARC), provided by NASA's Goddard Space Flight Center. This research made use of NASA's Astrophysics Data System. LM is supported by China Scholarship Council (CSC), grant number 201208440011. DA acknowledges support from the Royal Society.

\begin{appendix}
\section{results of the fit to the four segments with a disc component }
As shown in Table \ref{unlink2} the temperature of the disc is above 0.65 keV in the first and the last segments, where the mHz QPO is detected, while it is below 0.64 keV for the other two segments. The inner radius of the disc pegs at the lower boundary, 5.12$R_{g}$, for all segments, while the emissivity index of the line marginally decreases from about 2.7 in the first three segments to about 2.5 in the fourth one. The width of the {\sc gauss} component appears to decrease slightly in the first three segments, and then increases to 1.5$\pm$0.2 in the last segment. The two line models show different results of the energy of the line: The line is at about 6.9 keV in the first segment and then pegs at 6.97 keV in the other segments for the {\sc gauss} model, while in the {\sc kyrline} case the line is between 6.6 keV and 6.7 keV in the first three segments, and then increases to $\sim$ 6.9 keV in the last segment.

The total unabsorbed flux and the flux of the {\sc nthcomp} evolve in a similar way as in the case where we used {\sc bbody+nthcomp} to fit the continuum: Both fluxes first increase and then remain more or less constant or just decrease slightly in the last segment. The {\sc diskbb} component is formally not required in the fits of segments 2 and 3, where there are no mHz QPO present in the data. For the flux of the iron line, in the case of {\sc gauss}, it is slightly lower in segments 2 and 3 where there is no mHz QPO, while in the {\sc kyrline} case the flux of the line is more or less constant in all segments. 

  The best fitting parameters to the spectra of the four segments using {\sc diskbb} to fit the soft component are given in Table \ref{link1} (linked parameters: $N_{H}$, $\Gamma$, $kT_{e}$ and $\theta$) and Table \ref{unlink2} (unlinked parameters), respectively.

\begin{table}
\center
\small
\caption{Fitting results of four segments of 4U 1636$-$53 using the continuum model {\sc diskbb+nthcomp}; unlinked parameters. We fit the four segments with two iron line models simultaneously (see text for more details); the $\chi^2_\nu$ $(\chi^2/dof)$ of the fit is 1.16 (1568/1348). }
\begin{tabular}{|c|c|c|c|c|c|}
\hline
\hline
Model comp    & Parameter        &   S1                   &S2                   &S3                &S4                      \\
\hline
{\sc diskbb} &  $kT$ (keV)        &  0.66$\pm 0.02      $& 0.60$\pm 0.03     $&  0.61$\pm 0.03        $&  0.67$\pm 0.02       $  \\
             &  Normalization   &  151.8$_{-5.6}^{+29.1} $&       $<  48.8     $&  $<  47.2   $&  90.4$_{-17.7}^{+38.2} $  \\
             &  Flux (10$^{-10}$) &5.1$\pm 1.8 $&    $<1.07$ &  $< 1.14$ &   3.2$\pm 2.2$ \\
{\sc nthcomp}&  Normalization   &  0.67$\pm 0.05       $&  0.83$\pm 0.01    $&  0.83$_{-0.06}^{+0.01}$&  0.73$\pm 0.06       $  \\
             &  Flux (10$^{-10}$)          &57.6$\pm 4.8 $&    66.5$\pm 6.2 $&    67.6$\pm 4.9 $&   63.5$\pm 6.1$ \\
{\sc kyrline}&$R_{\rm in}$ ($R_{G}$)&  5.12$_{-0}^{+0.18}    $&5.12$_{-0}^{+0.21}   $&  5.12$_{-0}^{+0.20}     $&  5.12$_{-0}^{+0.43}     $  \\
            &$E_{\rm line}$ (keV)&  6.65$\pm 0.07      $&  6.69$\pm 0.11     $&  6.62$\pm 0.09        $&  6.88$_{-0.1}^{+0.09}  $  \\
             &  $\alpha$        &  2.7$\pm 0.1         $&  2.7$\pm 0.2      $&  2.7$\pm 0.2         $&  2.5$\pm 0.2        $  \\
             &  Normalization (10$^{-3}$)  &  11.4$\pm 0.8       $&  10.5$\pm 0.8      $&  11.3$\pm 0.8         $&  10.6$\pm 1.3        $  \\
             &  Flux (10$^{-10}$)          &1.26$\pm 0.08 $&    1.17$\pm 0.09 $&    1.24$\pm 0.09 $&  1.2$\pm 0.1 $\\
             &  Total flux (10$^{-10}$)    &63.9$\pm 5.1 $&    67.7$\pm 6.6 $&    68.9$\pm 5.6$ &   68.0$\pm 6.4$ \\
\hline
{\sc diskbb} &  $kT$ (keV)        &  0.68$\pm 0.02      $&  0.61$\pm 0.03      $&  0.61$\pm 0.03       $&  0.69$\pm 0.03       $  \\
             &  Normalization   &  164.2$_{-11.0}^{+24.7}$&  $<  66.3         $&  $< 40.8         $&  106.6$_{-17.0}^{+33.0}$  \\
             &  Flux (10$^{-10}$)&6.2$\pm 1.9 $&   $< 1.53$ &    $<1.00$ &   4.2$\pm 2.8 $   \\
{\sc nthcomp}&  Normalization   &  0.64$\pm  0.03     $&  0.82$_{- 0.1}^{+0.02}$&  0.83$_{-0.1}^{+0.01}  $&  0.70$\pm 0.06       $  \\
             &  Flux (10$^{-10}$) &55.9$\pm 4.1 $&    66.1$\pm 6.5 $&    67.7$\pm 7.1 $&   62.0$\pm 6.2 $\\
{\sc gauss} &$E_{\rm line}$ (keV)&  6.88$_{-0.12}^{+0.09}$&  6.97$_{-0.10}^{  +0 } $&  6.97$_{- 0.05}^{+0}   $&  6.97$_{-0.06}^{+0}    $  \\
             &  $\sigma$ (keV)  &  1.5$\pm 0.1       $&  1.4$\pm 0.1       $&  1.33 $\pm 0.08       $&  1.5$\pm 0.2        $  \\
             &  Normalization (10$^{-3}$)  &  14.3$\pm 2.1       $&  11.0$\pm 1.2       $&  11.1$\pm 1.0        $&  13.0$\pm 2.5        $  \\
             &  Flux (10$^{-10}$)&1.6$\pm 0.2 $&    1.2$\pm 0.1 $&    1.3$\pm 0.1$ &   1.5$\pm 0.3$ \\
             &  Total flux (10$^{-10}$)    & 63.7$\pm 4.6 $&    67.6$\pm 7.1 $&    69.0$\pm  7.6 $&   67.7$\pm 6.8$ \\             
\hline
\hline
\end{tabular}
\medskip
\\
\label{unlink2}
\end{table}

\end{appendix}

\clearpage

\bibliographystyle{mn}
\bibliography{biblio}

\label{lastpage}

\end{document}